\documentclass[%
final,
conference %
comsoc,%
letterpaper,%
oneside,%
twocolumn,%
nofonttune,%
]{IEEEtran}%
\PassOptionsToPackage{usenames,dvipsnames,svgnames,x11names,table,prologue}{xcolor}%
\PassOptionsToPackage{hyphens}{url}%
\PassOptionsToPackage{nomessages}{fp}%

\ifCLASSINFOpdf
  \usepackage[pdftex]{graphicx}
\else
  \usepackage[dvips]{graphicx}
\fi

\ifCLASSOPTIONcompsoc
   \usepackage[caption=false,font=normalsize,labelfont=sf,textfont=sf]{subfig}
\else
   \usepackage[caption=false,font=footnotesize]{subfig}
\fi

\usepackage{stfloats}

\usepackage[english]{babel}%
\usepackage[utf8]{inputenc}%
\usepackage[babel,style=english]{csquotes}%
\usepackage{hyphsubst}%
\usepackage[%
	activate={true,%
	nocompatibility},%
	final,%
	tracking=true,%
	kerning=true,%
	spacing=true,%
	factor=1100,%
	stretch=10,%
	shrink=10%
]{microtype}%
\usepackage{setspace}%
\usepackage{xcolor}%
\definecolor{todonotecol}{RGB}{250,0,0}%
\usepackage{xparse}
\usepackage{url}%
\usepackage[%
	acronym,%
	nopostdot,%
	seeautonumberlist,%
	shortcuts,%
	section=chapter,%
	toc,%
]{glossaries}%
\glsaddkey*{url}%
{}%
{\glsentryurl}%
{\Glsentryurl}%
{\glsurl}%
{\Glsurl}%
{\GLSurl}%
\glsaddkey*{longpltwo}%
{}%
{\glsentrylongpltwo}%
{\Glsentrylongpltwo}%
{\glslongpltwo}%
{\Glslongpltwo}%
{\GLSlongpltwo}%
\newglossaryentry{}{%
	name={},%
	plural={},%
	description={},%
}%
\newacronym{it}{IT}{information technology}
\newacronym{ot}{OT}{operational technology}
\newacronym{opcua}{OPC UA}{Open Platform Communications Unified Architecture}
\newacronym{tsn}{TSN}{Time-Sensitive Networking}
\newacronym{scada}{SCADA}{supervisory control and data acquisition}
\newacronym{vpc}{VPC}{Virtual Process Controller}
\newacronym{plc}{PLC}{Programmable Logic Controller}
\newacronym{udp}{UDP}{User Datagram Protocol}
\newacronym{pubsub}{PubSub}{Publish Subscribe}
\newacronym{vm}{VM}{virtual machine}
\newacronym{pc}{PC}{personal computer}
\newacronym{ptp}{PTP}{Precision Time Protocol}
\newacronym{gptp}{gPTP}{generalized Precision Time Protocol}
\newacronym{ntp}{NTP}{Network Time Protocol}
\newacronym{splc}{Soft-PLC}{Software-PLC}
\newacronym{rte}{RTE}{Real-Time Ethernet}
\newacronym{osi}{OSI}{Open Systems Interconnection}
\newacronym{mac}{MAC}{Media Access Control}
\newacronym{e2e}{E2E}{end-to-end}
\newacronym{tcp}{TCP}{Transmission Control Protocol}
\newacronym{hmi}{HMI}{Human-Machine Interface}
\newacronym{kpi}{KPI}{Key Performance Indicator}
\newacronym{mes}{MES}{Manufacturing Execution System}
\newacronym{rami4.0}{RAMI 4.0}{Reference Architectural Model Industrie 4.0}
\newacronym{tacnet4.0}{TACNET 4.0}{Tactile Internet 4.0}
\newacronym{3gpp}{3GPP}{3rd Generation Partnership Project}
\newacronym{5gppp}{5GPPP}{5G Infrastructure Public Private Partnership}
\newacronym{itu}{ITU}{International Telecommunication Union}
\newacronym{ngnm}{NGNM}{Next Generation Mobile Networks}
\newacronym{agv}{AGV}{automated guided vehicle}
\newacronym{v2v}{V2V}{vehicle-to-vehicle}
\newacronym{v2x}{V2X}{vehicle-to-infrastructure}
\newacronym{d2x}{D2X}{device-to-x}
\newacronym{qos}{QoS}{quality of service}
\newacronym{noa}{NOA}{NAMUR Open Architecture}
\newacronym{dcs}{DCS}{distributed control system}
\newacronym{m2m}{M2M}{machine-to-machine}
\newacronym{sdn}{SDN}{software-defined networking}
\newacronym{erp}{ERP}{enterprise resource planning}
\newacronym{ip}{IP}{Internet Protocol}
\newacronym{lte}{LTE}{Long Term Evolution}
\newacronym{5g}{5G}{5th generation wireless communication systems}
\newacronym{harq}{HARQ}{Hybrid Automatic Repeat Request}
\newacronym{rat}{RAT}{radio access technology}
\newacronym{mc}{MC}{Multi-Connectivity}
\newacronym{fbmc}{FBMC}{Filter Bank Multicarrier}
\newacronym{gfdm}{GFDM}{Generalized Frequency Division Multiplexing}
\newacronym{sla}{SLA}{service-level agreement}
\newacronym{5qi}{5QI}{5G QoS Indicator}
\newacronym{bmbf}{BMBF}{German Federal Ministry of Education and Research}
\newacronym{nrt}{NRT}{non real-time}
\newacronym{irt}{IRT}{isochronous real-time}
\newacronym{rt}{RT}{real-time}
\newacronym{ar}{AR}{augmented reality}
\newacronym{wlan}{WLAN}{Wireless Local Area Network}
\newacronym{cpu}{CPU}{central processing unit}
\newacronym{ram}{RAM}{random-access memory}
\newacronym{ue}{UE}{user equipment}
\newacronym{ran}{RAN}{radio access network}
\newacronym{bs}{BS}{base station}
\newacronym{cn}{CN}{core network}
\newacronym{epc}{EPC}{evolved packet core}
\newacronym{mec}{MEC}{mobile edge cloud}
\newacronym{rtt}{RTT}{Round Trip Time}
\newacronym{vpn}{VPN}{virtual private network}
\newacronym{vpf}{VPF}{Virtual Process Control Function}
\newacronym{vdi}{VDI}{Association of German Engineers}
\newacronym{ipc}{IPC}{industrial pc}
\newacronym{rtos}{RTOS}{real-time operating system}
\newacronym{mano}{VPC M\&O}{VPC Manager \& Orchestrator}
\newacronym{i/o}{I/O}{Input / Output}
\newacronym{fbd}{FBD}{Function Block Diagram}
\newacronym{ldd}{LDD}{Ladder Logic Diagram}
\newacronym{sfc}{SFC}{Sequential Function Chart}
\newacronym{il}{IL}{Instruction List}
\newacronym{st}{ST}{Structured Text}
\newacronym{os}{OS}{Operating System}
\newacronym{mqtt}{MQTT}{message queuing telemetry transport }
\newacronym{amqp}{AMQP}{advanced message queuing protocol}
\newacronym{api}{API}{application programming interface}
\newacronym{embb}{eMBB}{enhanced mobile broadband}
\newacronym{iot}{IoT}{Internet of Things}
\newacronym{urllc}{URLLC}{ultra-reliable low-latency communication}
\newacronym{mmtc}{mMTC}{massive machine-type communications}
\newacronym{iiot}{IIoT}{industrial IoT}
\newacronym{iiot2}{IIoT}{industrial Internet of Things}
\newacronym{npn}{NPN}{non-public network}
\newacronym{prb}{PRB}{Physical Ressource Block}
\newacronym{gnb}{gNB}{5G base station}%
\newacronym{5gs}{5GS}{5G system}
\newacronym{5gc}{5GC}{5G core network}
\newacronym{oai}{OAI}{OpenAirInterface}
\newacronym{enb}{eNB}{4G base station}
\newacronym{sdr}{SDR}{Software Defined Radio}
\newacronym{usrp}{USRP}{Universal Software Radio Peripheral}
\newacronym{mme}{MME}{Mobility Management Entity}
\newacronym{hss}{HSS}{Home Subscriber Server}
\newacronym{pdn}{PDN}{Packet Data Network}
\newacronym{sgw}{SGW}{Serving Gateway}
\newacronym{gtp}{GTP}{GPRS Tunneling Protocol}
\newacronym{mimo}{MIMO}{Multiple Input Multiple Output}
\newacronym{fpga}{FPGA}{Field Programmable Gate Array}
\newacronym{pss}{PSS}{primary synchronization signal}
\newacronym{sss}{SSS}{secondary synchronization signal}
\newacronym{sfn}{SFN}{system frame number}
\newacronym{mib}{MIB}{master information block}
\newacronym{pbch}{PBCH}{Physical Broadcast Channel}
\newacronym{ism}{ISM}{Industrial, Scientific and Medical}
\newacronym{e-utra}{E-UTRA}{evolved UMTS Terrestrial Radio Access}
\newacronym{bldc}{BLDC}{brushless DC}
\newacronym{dl}{DL}{Downlink}
\newacronym{ds-tt}{DS-TT}{Device-Side TSN Translator}
\newacronym{nw-tt}{NW-TT}{Network-Side TSN Translator}
\newacronym{upf}{UPF}{User Plane Function}
\newacronym{tsi}{TSi}{ingress timestamping}
\newacronym{tse}{TSe}{egress timestamping}
\newacronym{gm}{GM}{grandmaster}
\newacronym{sn}{SN}{sequence number}
\newacronym{rbis}{RBIS}{Reference Broadcast Infrastructure Synchronization}
\newacronym{cococo}{CoCoCo}{communication-control codesign}
\newacronym{3c}{3C}{joint design of communications, control, and computing}
\newacronym{cots}{COTS}{Commercial Off-The-Shelf}
\newacronym{ie}{IE}{industrial Ethernet}
\newacronym{ap}{AP}{access point}
\newacronym{ofdm}{OFDM}{Orthogonal Frequency-Division Multiplexing}
\newacronym{csma}{CSMA-CA}{carrier sense multiple access with collision avoidance}
\newacronym{dcf}{DCF}{distributed coordination function}
\newacronym{pcf}{PCF}{point coordination function}
\newacronym{hcf}{HCF}{hybrid coordination function}
\newacronym{edca}{EDCA}{enhanced distributed channel access}
\newacronym{hcca}{HCCA}{\gls{hcf} controlled channel access}
\newacronym{tdma}{TDMA}{time division multiple access}
\newacronym{ofdma}{OFDMA}{Orthogonal Frequency-Division Multiple Access}
\newacronym{mumimo}{MU-MIMO}{multi-user \gls{mimo}}
\newacronym{bi}{BI}{beacon interval}
\newacronym{ssid}{SSID}{service set identifier}

\glsdisablehyper%
\usepackage[%
	backend=biber,%
	style=ieee,%
	isbn=false,%
	hyperref=true,%
	maxbibnames=99,%
	sorting=none,%
	natbib=true,%
	language=english,%
	defernumbers=true,%
    hyperref=false
	]{biblatex}%
\DeclareFieldFormat{sentencecase}{\csname bbx@colon@search\endcsname#1}%

\usepackage{nameref}
\usepackage{soul}
\usepackage{flushend}
\usepackage{textcomp}
\usepackage{calc}
\usepackage{xkeyval}
\usepackage{multirow}
\usepackage{tabulary}%
\usepackage{makecell}%
\usepackage{pgfplots}
\usepackage{tikz}
\usepackage{textcomp} 
\usepackage{verbatim}

\newcommand{\nl}{\par\noindent} %

\newcommand{\mytilde}{{\raise.17ex\hbox{$\scriptstyle\mathtt{\sim}$}}}

\newlength\textheighttemp%
\newlength\textwidthtemp%
\newlength\textheightstd%
\newlength\textwidthstd%
\newlength\textheightold%
\newlength\textwidthold%
\newlength\tempheight%
\newlength\tempwidth%
\SetProtrusion{encoding={*},family={bch},series={*},size={6,7}}
              {1={ ,750},2={ ,500},3={ ,500},4={ ,500},5={ ,500},
               6={ ,500},7={ ,600},8={ ,500},9={ ,500},0={ ,500}}
\SetExtraKerning[unit=space]
    {encoding={*}, family={bch}, series={*}, size={footnotesize,small,normalsize}}
    {\textendash={400,400}, %
     "28={ ,150}, %
     "29={150, }, %
     \textquotedblleft={ ,150}, %
     \textquotedblright={150, }} %
\SetTracking{encoding={*}, shape=sc}{40}
\makeatletter
\let\blx@rerun@biber\relax
\makeatother
\usepackage{listings}
\usepackage[dvipsnames]{xcolor}
\usepackage{balance}
\usepackage{tikz} \usetikzlibrary{calc, arrows.meta, intersections, patterns, positioning, shapes.misc, fadings, through,decorations.pathreplacing}

\definecolor{ColorOne}{named}{MidnightBlue}
\definecolor{ColorTwo}{named}{Dandelion}
\definecolor{ColorThree}{named}{Plum}

\usepackage{algorithm}
\usepackage{algpseudocode}
\usepackage{tabularx}
\usepackage{newtxmath}
\usepackage{subcaption}
\usepackage{booktabs}
\pgfplotsset{
	grid style = {
		line width = 0.1pt
	}
}
\newcommand{\disablewr}[1]{#1}%
\newcommand{\newcommanddisw}[3]{\newcommand{#1}[1]{\disablewr{\textcolor{#2}{#3}}}}%
\renewcommand{\disablewr}[1]{}%
\definecolor{todocol}{named}{red}
\newcommanddisw{\todo}{todocol}{ToDo: #1}%
\definecolor{migucol}{named}{purple}%
\newcommanddisw{\migucom}{migucol}{{@}comment: #1}%
\newcommanddisw{\miguhigh}{migucol}{#1}%
\usepackage{siunitx}
\usepackage{xspace}

\newcommand{\TempDisplayPreparation}{\disablewr{%
		\section{Draft-State: Comment Color Code}\noindent%
		\todo{Comments: ToDos}\nl%
		\migucom{Comments: Michael Gundall}\nl%

}}%
\begin{document}
\title{%
	FLEX: Joint UL/DL and QoS-Aware Scheduling for Dynamic TDD in Industrial 5G and Beyond
	\thanks{This research was supported by the German Federal Ministry of Research, Technology and Space (BMFTR) within the projects Open6GHub+ and 6GTerafactory under grant numbers 16KIS2402K and 16KISK186.
		The responsibility for this publication lies with the authors.
		This is a preprint of a work accepted but not yet published at the 2026 IEEE International Conference on Communications.
		Please cite as: L. Kleinberger, M. Gundall and H. D. Schotten: “FLEX: Joint UL/DL and QoS-Aware Scheduling for Dynamic TDD in Industrial 5G and Beyond”. In: 2026 IEEE International Conference on Communications (ICC), IEEE, 2026.}
}
\author{%
\IEEEauthorblockN{%
    Leonard Kleinberger\IEEEauthorrefmark{1}, %
    Michael Gundall\IEEEauthorrefmark{1}, %
    and Hans D. Schotten\IEEEauthorrefmark{1}\IEEEauthorrefmark{3} %
    \\%
}%
\IEEEauthorblockA{%
    \IEEEauthorrefmark{1}German Research Center for Artificial Intelligence GmbH (DFKI), Kaiserslautern, Germany \\%
    \IEEEauthorrefmark{3}Department of Electrical and Computer Engineering,  RPTU University Kaiserslautern-Landau, Kaiserslautern, Germany %
	\\%
    Email: %
        \{leonard.kleinberger, michael.gundall, hans\_dieter.schotten%
        \}@dfki.de%
        \\%
}%
}%

\maketitle
\begin{abstract}%
	Industrial 5G deployments using Time Division Duplex (TDD) networks face a critical challenge: existing schedulers rely on static configuration of Uplink (UL) to Downlink (DL) resource ratios, failing to adapt to dynamic asymmetric traffic demands.
	This limitation is particularly problematic in Industry 4.0 scenarios where traffic patterns exhibit significant asymmetry between directions and heterogeneous Quality of Service (QoS) requirements.
	We present FLEX, a novel QoS-aware scheduler that dynamically adjusts the UL/DL ratio in flexible TDD slots while respecting diverse QoS requirements.
	FLEX introduces DL buffer state estimation to prevent starvation of high-priority DL traffic, exploiting the deterministic nature of industrial traffic patterns for accurate predictions.
	Through extensive simulations of industrial scenarios using 5G LENA and ns-3, we demonstrate that FLEX achieves similar throughput compared to established scheduling while correctly enforcing QoS priorities in both traffic directions.
	For deterministic traffic patterns, FLEX maintains minimal latency overhead (less than 1 slot duration), making it particularly suitable for industrial automation applications.
\end{abstract}

\IEEEpeerreviewmaketitle
\tikzstyle{descript} = [text = black,align=center, minimum height=1.8cm, outer sep=0pt,font = \footnotesize]
\tikzstyle{activity} =[align=center,outer sep=1pt]

\section{Introduction}%
\label{sec:Introduction}

The Fifth Generation (5G) of mobile networks represents a transformative shift in wireless communication, extending beyond traditional mobile broadband to enable critical industrial applications.
In the context of Industry 4.0, smart factories integrate sensors, actuators, and control systems into data-driven environments, requiring high-performance communication systems as the backbone of all operations.
To fulfill the stringent requirements of industrial networks and to allow the deployment of many adjacent cells, operators have turned towards Time Division Duplex Mode (TDD)~\cite{adamuzhinojosa2025empiricalanalysis5gtdd}.

Unlike Frequency Division Duplex (FDD), where Uplink (UL) and Downlink (DL) operate simultaneously on separate frequency bands, TDD shares the same frequency for both directions.
This provides the flexibility needed to address asymmetric traffic demands while ensuring efficient spectrum usage.
To adapt the TDD system to different scenarios, 5G introduces dynamic TDD, which enables the declaration of time slots as UL, DL or flexible and allows changes to the TDD pattern during operation~\cite{5G_advanced}.
Flexible slots are especially interesting, because they can carry traffic in either direction without static allocation.

In practice however, 5G TDD schedulers rarely exploit the full potential of dynamic slot allocation~\cite{10.1145/3696395}.
Instead, they typically rely on initial static configuration determined by the operator, requiring complex pre-deployment computations.
Even when flexible slots are configured, simple rules are commonly applied that generally favour either UL or DL, leading to resource underutilization and quality of service violations in asymmetric scenarios.

This paper introduces FLEX, a flexible QoS-aware scheduler designed to dynamically adjust the UL/DL ratio in TDD flexible slots while doing fair scheduling between UL and DL.
The key innovation of FLEX is a DL buffer state estimation during UL scheduling time, which allows the system to reserve resources for high-priority DL traffic before the UL scheduler makes irrevocable allocations.
For industrial scenarios with deterministic traffic patterns, FLEX further exploits traffic predictability to minimize unused resources in guard symbols in order to maximize global performance.

\textbf{Contributions.}
This paper makes the following contributions:
\begin{itemize}
	\item A novel scheduler design that dynamically adjusts UL/DL ratios in flexible TDD slots while maintaining QoS fairness between both traffic directions.
	\item A buffer state estimation technique that exploits deterministic characteristics to avoid asymmetric timing constraints between both traffic directions.
	\item A guard symbol avoidance mechanism that reduces the number of necessary guard symbols to maximize cell performance.
\end{itemize}

\textbf{Paper Structure.}
Section~\ref{sec:Related Work} discusses related work on 5G scheduling.
Section~\ref{sec:System Model} presents the system model and key challenges.
Section~\ref{sec:FLEX Design} details the FLEX scheduler design.
Section~\ref{sec:Evaluation} describes the evaluation methodology and presents simulation results.
Finally, Section~\ref{sec:Conclusion} concludes the paper.

\section{Related Work}%
\label{sec:Related Work}
5G scheduling has been extensively studied in the literature, with most work focusing on either UL or DL independently~\cite{9773317}.
The core of each scheduling procedure is a priority function, which is used to determine the prioritized flows.
Classical approaches to scheduling include Proportional Fair (PF) and Maximum Rate (MR).

For industrial applications, several QoS-aware schedulers have been proposed~\cite{9773317}.
These typically prioritize traffic based on 5G QoS Identifiers (5QI), which encode resource type, priority level, packet delay budget, and packet error rate requirements.
However, existing QoS schedulers operate independently in UL and DL directions without coordination, leading to resource conflicts in flexible TDD slots.

Some work has also been done to adjust the semi-static TDD configuration during operation~\cite{8944281,9127428}.
However, these approaches work by adjusting the TDD pattern over larger durations, thereby not allowing them to adjust to fast switching between UL and DL heavy transmission.
The challenge of UL/DL asymmetry in TDD systems has been recognized, yet proposed solutions typically rely on traffic prediction without addressing the fundamental timing constraint that gives UL scheduling precedence over DL.

Our work differs fundamentally by introducing buffer estimation, which can be used to equalize the different timing constraints between UL and DL.
This buffer estimation therefore enables fast switching of transmission directions using only flexible slots.

\section{System Model and Key Challenges}%
\label{sec:System Model}
\subsection{5G TDD Architecture}
We consider a single-cell 5G TDD system with one gNB serving multiple UEs.
The system operates with flexible slot configuration, where each slot can dynamically carry UL or DL traffic, or a mixture of both.
In the time domain, the basic unit is a slot consisting of $14$ OFDM symbols.
For the frequency domain, resources are organized as Resource Block Groups (RBGs).
The scheduler operates on a two-dimensional resource grid of RBGs versus symbols~\cite{ETSI_TS_138_211}.
All allocations in the resource grid have to be done in the TDMA scheme.

\begin{figure}[t]
	\begin{center}
		\includegraphics[width=\columnwidth]{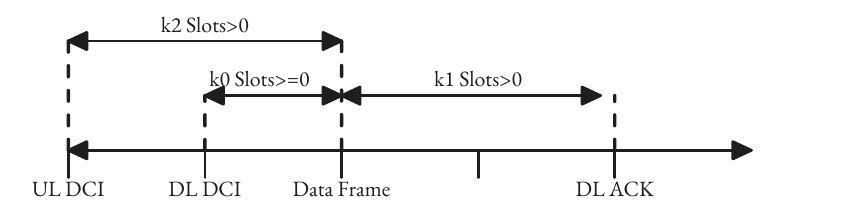}
	\end{center}
	\caption[]{The timeline for a data transmission in 5G systems. Control information is sent via DCI before the actual data transmission. The UL DCI Deadline is before the DL DCI Deadline.}
	\label{fig:timeline}
\end{figure}

For scheduling transmissions, several delays have to be taken into account between grant transmission and the actual data transfer as depicted in Figure~\ref{fig:timeline}.
The grant for DL transmissions has to be sent $k0$ slots before the data transmission~\cite{ETSI_TS_138_214}.
The gNB is both the scheduler and the DL sender, so $k0$ is usually set to $0$.
For UL transmissions, the $k2$ parameter defines the minimum number of slots between UL grant transmission and actual UL data transmission.
This delay accounts for UE processing time, receive-to-transmit mode switching and propagation delay.
$k2$ is always larger than $k0$.
Typically, $k2=2$ slots which is approximately \SI{1}{ms} for numerology $1$.

\subsection{UL Scheduling Procedure}
When a UE has UL data to transmit, it sends a Scheduling Request (SR) via the Random Access Channel.
Upon reception of the SR, the gNB becomes aware of pending data but lacks information about data amount and QoS requirements.
The gNB therefore schedules the UE for transmission of a Buffer Status Report (BSR).
The resource allocation for the BSR is transmitted via UL Downlink Control Information (UL DCI).
After $k2$ slots, the BSR transmission occurs, providing the gNB with accurate buffer occupancy and QoS information.
The gNB then schedules the actual data transmission, which occurs after an additional $k2$ slots~\cite{ETSI_TS_138_321}.

The BSR reports three logical queues: (1) status information (ACK/NACK), (2) retransmission queue, and (3) new transmission queue.
Data is prioritized in this order when allocating UL resources. %

DL data does not have to abide by this procedure, because the DL buffers are located at the gNB.
Therefore, the scheduler, which is also located at the gNB, always has full information about the DL buffer status.

\subsection{The DL Starvation Problem}
The asymmetric timing constraints create a fundamental scheduling challenge.
For a given slot $t$:
\begin{itemize}
	\item Before scheduling UL data, the information about the UL buffer status must be gathered with a scheduled BSR.
	      This is a priority inversion problem, because the BSR usually does not have a high priority due to its small size.
	\item UL scheduling decisions must be finalized at slot $t-k2$ ($k2$ slots ahead)
	\item DL scheduling decisions can be finalized at slot $t$ (same slot)
\end{itemize}

Multiple solutions exist to deal with these constraints.
By defining a fixed UL/DL ratio, the different deadlines between UL and DL do not have any effect on scheduling fairness, because UL and DL scheduling can be done independently.
A different solution is to have the UL scheduler operate first on all available resources and the DL scheduler operating later on the remaining resources.
This solution inherently favours UL traffic, because even low priority UL traffic can starve high priority DL traffic.
A more advanced solution to this problem is doing joint UL and DL Scheduling at UL scheduling time.
This solution is more fair than separate UL-DL scheduling but DL scheduling experiences an additional and unnecessary $k2$ delay latency and the BSR priority inversion problem still persists.

\subsection{QoS Requirements}

In 5G traffic is classified using 5G QoS Identifiers (5QI)~\cite{ETSI_TS_123_203}, which define:
\begin{itemize}
	\item \textbf{Resource Type}: Guaranteed bit rate (GBR) or non-GBR
	\item \textbf{Priority}: Integer from $1$ (highest) to $127$ (lowest)
	\item \textbf{Packet Delay Budget}: Maximum tolerable end-to-end delay
	\item \textbf{Packet Error Rate}: Maximum acceptable loss rate
\end{itemize}

The assignment of 5QI to traffic is done by the gNB and the UE based on predefined traffic filters.
The scheduler must respect these requirements for both UL and DL traffic while allocating resources fairly between directions.

\section{FLEX Scheduler Design}%
\label{sec:FLEX Design}
FLEX addresses the DL starvation problem through four key phases.
\begin{enumerate}
	\item \textbf{Bidirectional Traffic Measurement}: Measure the traffic ingress for each flow
	\item \textbf{Buffer State Estimation}: Use the traffic measurement to compensate the $k2$ scheduling delay and to avoid latency introduced by BSR in UL.
	\item \textbf{Joint UL/DL Scheduling}: Restrict UL scheduler from consuming resources needed for high-priority DL traffic.
	\item \textbf{DL Scheduling Re-evaluation}: Re-evaluate the DL scheduling decisions done previously to react to unexpected changes.
\end{enumerate}

\subsection{Bidirectional Traffic Measurement}
FLEX maintains historical buffer status $Q_{i,f}(t)$ and resource allocations $A_{i,f}(t)$ for each logical channel (LC) $i$ for traffic flow $f$ over time in ring buffers.
For UL, an accurate buffer status is only available after the reception of the BSR.
When BSRs are not available, missing buffer states are reconstructed using:
\begin{equation}
	Q_{i,f}(t) = \max(Q_{i,f}(t-1) - A_{i,f}(t-1), 0)
	\label{eq:buffer_reconstruction}
\end{equation}
This reconstruction is equal to estimating the ingress as $0$.
Based on the historical buffer status and the resource allocations, the ingress $I_{i,f}$ is calculated as
\begin{equation}
	I_{i,f}(t) = Q_{i,f}(t) - Q_{i,f}(t-1) + \min(A_{i,f}(t-1), Q_{i,f}(t-1))
	\label{eq:ingress}
\end{equation}
For industrial devices, the traffic pattern is assumed to be semi-static.
This means that message sizes are expected to be fixed and the message interval to be uniformly distributed around a known value.
The distribution of message intervals simulates different clock rates at different devices.
From historical ingress values, FLEX computes for each flow:
\begin{itemize}
	\item \textbf{Burst Size} ($B_{i,f}$): Mean of all non-zero ingress values
	\item \textbf{Burst Interval} ($T_{i,f}$): Mean number of slots between bursts
	\item \textbf{Prediction Confidence} (CV = $\sigma/\mu$):  Coefficient of Variation of Burst Interval and Burst Size respectively
\end{itemize}
For UL Traffic, the scheduler recognizes packet arrivals only after the BSR has been received and processed.
However, because the time of transmission for the BSR is known to the gNB, the packet arrival can be backlogged to the time the BSR has been sent.
The CV indicates whether traffic patterns are stable enough to perform Buffer State Estimation.
If CV indicates low variability (deterministic traffic), FLEX predicts the next burst arrival time and size.

\subsection{Buffer State Estimation}
The key innovation of FLEX is estimating buffer states at UL scheduling time (slot $t-k2$) to enable coordinated resource allocation.
This estimation addresses two critical challenges: (1) compensating for the $k2$ timing asymmetry between UL and DL scheduling, and (2) avoiding the additional latency introduced by BSR requests in UL.

\textbf{DL Buffer State Estimation.}
For DL traffic, the actual buffer state $Q_{i,f}(t)$ is known at slot $t$ but unknown at UL scheduling time $t-k2$.
FLEX estimates the future DL buffer state using the measured traffic characteristics.
The ingress for the semi-static traffic patterns is modelled using a step function where the step width is defined by the estimated burst interval and the step height is defined by the estimated burst size.
Modelling the traffic pattern using the step function allows predicting the buffer ingress iteratively for each future frame without modelling partial frame arrivals.

If the CV indicates high variability (indeterministic traffic), the DL Buffer State Estimation is skipped.
Skipping the DL Buffer State Estimation for indeterministic traffic ensures that the estimated traffic demand does not deviate from the true traffic demand.
This way, the number of allocations for traffic demands, which do not exist, can be reduced.
DL buffers with indeterministic traffic can still be used for restricting the UL scheduler resource consumption thereby limiting the DL starvation problem.
However, because the $k2$ delay cannot be compensated, an additional latency of $k2$ slots for DL cannot be avoided.

\textbf{UL Buffer State Estimation.}
For UL traffic, FLEX faces an additional challenge: buffer states are located at the UE and only known after BSR reception.
On one hand, BSRs have to be sent regularly to accurately inform the gNB about incoming traffic.
On the other hand, too many BSRs can easily lead to network congestion by using resources that would be better used for data traffic.
For this reason, BSRs should be able to be prioritized over traffic, but only if they inform about traffic, which cannot be inferred otherwise.

For semi-static industrial patterns, the inference of the traffic pattern is generally possible from the bidirectional traffic measurement.
If the CV indicates low variability (deterministic traffic), the probability of BSR being received at regular intervals and showing similar ingress is high.
For BSRs only showing similar ingress, the BSR transmission is not of high priority.
Instead, the buffer state can be directly estimated from the previously measured traffic patterns and resource allocations can be done accordingly.
A side effect of directly estimating the ingress from just the indicated existence of regular BSRs is that the UL scheduling procedure can potentially be reduced by one roundtrip.
This also reduces the latency by a minimum of $k2$ slots.

However, if the BSR interval does not fit into the previously measured pattern or the traffic pattern is indeterministic, the ingress cannot be assumed from the indicated existence of the BSR and the BSR cannot be skipped.
As a result, the irregular BSRs do require a high priority and the latency for indeterministic traffic cannot be reduced.

\subsection{Joint UL/DL Scheduling Procedure}
With buffer states estimated for both UL and DL traffic, FLEX performs coordinated resource allocation that respects QoS priorities across both directions.
The joint scheduling procedure operates in two steps: first, calculating priorities for all flows in both directions; second, selecting the optimal resource allocation strategy for the current slot.
This two-stage approach enables FLEX to balance QoS enforcement with efficient resource utilization.

\begin{figure*}[t]
	\begin{center}
		\includegraphics[width=1.5\columnwidth]{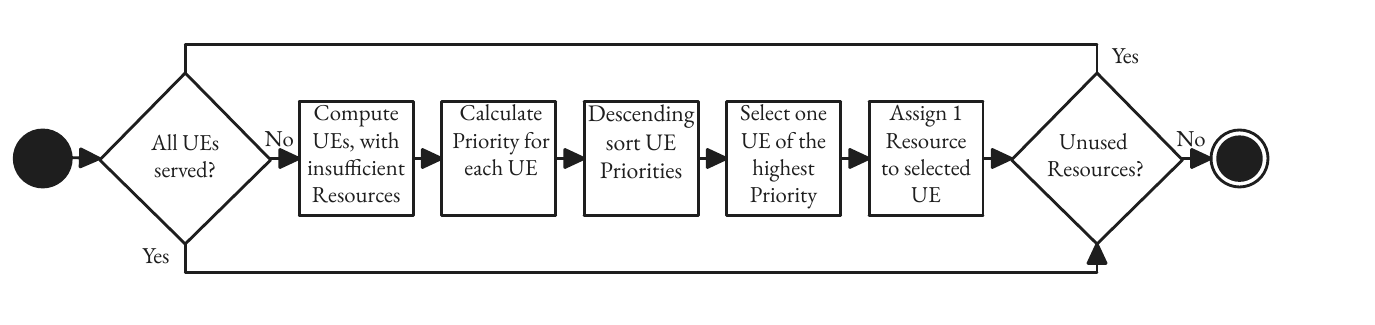}
	\end{center}
	\caption[]{Activity diagram for the scheduling procedure. Resources are assigned one-by-one to the UE with the highest priority.}
	\label{fig:flex_algorithm}
\end{figure*}

\textbf{Priority Function.}
To determine the priority for each dataflow, FLEX calculates a priority for each dataflow in each traffic direction.
To do this, FLEX propagates the 5QI QoS priority through the stack to the mac scheduler.
Note that in 5QI a lower QoS priority value translates to higher priority.
FLEX uses the Priority function:
\begin{equation}
	w_{i,f}(t) = \frac{1}{\text{5QI\_Priority}_{i,f}} \times \frac{r_{i,f}(t)}{R_{i,f}(t)}
	\label{eq:priority}
\end{equation}
where $r_{i,f}(t)$ is the instantaneous achievable data rate of LC $i$ at flow $f$ and $R_{i,f}(t)$ is the past average data rate.
The instantaneous achievable data rate directly depends on the current buffer status and the current achievable MCS.
The past average data rate can easily be estimated from the ring buffers that are used for the Bidirectional Traffic Measurement phase.
In short, the selected priority function combines the QoS priority with a Proportional Fair term to balance QoS requirements with fairness.
The scheduling procedure is detailed in Figure~\ref{fig:flex_algorithm}.
\textbf{Strategy Selection.}
For each flexible slot, FLEX evaluates three scheduling strategies:
\begin{enumerate}
	\item \textbf{UL-only}: Allocate all symbols to UL (requires guard symbols for DL-to-UL transition)
	\item \textbf{DL-only}: Allocate all symbols to DL (requires no guard symbols)
	\item \textbf{Mixed}: Mix UL and DL within the slot (requires guard symbols)
\end{enumerate}

For each scheduling strategy, FLEX evaluates the reward as the sum of the priorities of all successfully scheduled transmissions.
The scheduling strategy with the highest reward is assumed to provide the best cell-wide performance.
By choosing between the three scheduling strategies, unnecessary resource usage through guard symbols and old retransmissions can be minimized.

The joint UL/DL scheduling phase is run directly at UL deadline.
Therefore, UL transmissions do have to be sent directly to the UE via DCI.
The DL deadline is typically $k2$ slots later, at which time the DL scheduling re-evaluation phase will be run.

\subsection{DL Scheduling Re-evaluation}
The deadline for the scheduling of DL resources is $k2$ slots after the deadline for scheduling of UL resources.
As a result, the DL allocations do not have to be fixed during the joint UL/DL scheduling procedure.
Instead, they are only used for restricting the UL scheduler from occupying resources, which would be better allocated to DL transmissions.

Before the deadline for DL scheduling but after the deadline for UL scheduling, new high priority DL transmissions or HARQ might arrive.
The DL scheduling re-evaluation phase is run at DL-scheduling time to prevent latency increases when unexpected, high-priority DL transmissions arrive between the UL and DL scheduling deadlines.

To accommodate for unexpected DL transmission between DL scheduling deadline and UL scheduling deadline, previously reserved resources for DL can be reassigned.
In order to determine whether a reassignment is necessary, the reward for previously assigned DL transmissions is compared with the updated set of DL transmissions.
In case a reassignment was performed, all currently planned, but not yet fixed, DL allocations are recalculated.
At most, this will lead to the recalculation of $k2$ slots, which is acceptable in terms of performance.

After the DL scheduling re-evaluation phase, the DL allocations can be sent to the UE.

\section{Evaluation}%
\label{sec:Evaluation}
\subsection{Simulation Setup}
\begin{table*}[t]
	\caption{Evaluation Scenario Parameters}\label{tab:Parameters}
	\begin{center}
		\footnotesize
		\begin{tabular}[c]{l||l|l|l}
			\hline
			\multicolumn{1}{c||}{\textbf{Parameter}}                 &
			\multicolumn{1}{c|}{\textbf{Scenario 1: Motion Control}} &
			\multicolumn{1}{c|}{\textbf{Scenario 2: Mobile Robots}}  &
			\multicolumn{1}{c}{\textbf{Scenario 3: Heterogeneous QoS}}                                                                                                   \\
			\hline
			Service Area                                             & \qtyproduct{50 x 10 x 10}{m} & \qtyproduct{1 x 1 x 0.01}{km}    & \qtyproduct{30 x 30 x 10}{m}    \\
			\hline
			UE Speed                                                 & \SI{36}{km/h}                & \SI{25}{km/h}                    & \SI{25}{km/h}                   \\
			\hline
			Flow 1 UEs                                               & \numrange{1}{20}             & \numrange{1}{80}                 & $4$                             \\
			Flow 1 pattern                                           & deterministic bidirectional  & semi-deterministic bidirectional & deterministic unidirectional UL \\
			Flow 1 message size                                      & \SI{50}{B}                   & \SI{145}{B}                      & \SI{50}{B}                      \\
			Flow 1 message interval                                  & \SI{500}{\mu s}              & $\SI{25}{ms} (\pm 25\%) $        & \SI{50}{\mu s}                  \\
			Flow 1 start time                                        & \SI{0}{s}                    & $\SI{0}{s}$                      & \SI{0}{s}                       \\
			\hline
			Flow 2 UEs                                               & -                            & -                                & $2$                             \\
			Flow 2 pattern                                           & -                            & -                                & deterministic unidirectional DL \\
			Flow 2 message size                                      & -                            & -                                & \SI{50}{B}                      \\
			Flow 2 message interval                                  & -                            & -                                & \SI{50}{\mu s}                  \\
			Flow 2 start time                                        & -                            & -                                & \SI{0}{s}                       \\
			\hline
			Flow 3 UEs                                               & -                            & -                                & $2$                             \\
			Flow 3 pattern                                           & -                            & -                                & deterministic unidirectional DL \\
			Flow 3 message size                                      & -                            & -                                & \SI{50}{B}                      \\
			Flow 3 message interval                                  & -                            & -                                & \SI{50}{\mu s}                  \\
			Flow 3 start time                                        & -                            & -                                & \SI{5}{s}                       \\
			\hline
		\end{tabular}
	\end{center}
\end{table*}
We evaluate the FLEX scheduler performance through three simulation scenarios in the 5G-LENA simulator~\cite{Patriciello2019} (ns-3 based~\cite{10.1145/2068897.2068948}).
At the time of development, 5G-LENA is the only platform supporting dynamic flexible slot scheduling.
However, TDMA is the only resource allocation method available for UL and DL in 5G-LENA at the time of writing.
The three simulation scenarios are based on industrial communication patterns described in \cite{ETSI_TS_122_104} and \cite{ETSI_TS_123_203}.

All scenarios use a single gNB with \SI{3.5}{GHz} frequency, \SI{20}{MHz} bandwidth, numerology $1$, and all-flexible TDD slot configuration (FFFFF pattern).

Channel propagation follows the NYUSIM Indoor Factory statistical model~\cite{7996792}.
The gNB is always placed at the centre of the service area.
UE mobility uses Brownian Motion with random direction and speed changes.
The $k2$ delay is set to $2$ slots and $2$ guard symbols are used when switching from DL to UL transmissions.

All traffic is implemented as IP-layer traffic, according to \cite{ETSI_TS_122_104} and is flowing from a UE to a server connected to the gNB or vice versa.
The IP header was measured to be \SI{20}{B}.
Notably, each UE has only one traffic flow with a single QoS class.
However, this traffic flow can be bidirectional.
Having only one QoS class per UE is important because allocations are sent to the UE without any identification of the actual scheduled traffic flow.
We compare FLEX against three baseline schedulers already implemented in 5G-LENA: Proportional Fair~\cite{4064107}, Maximum Rate~\cite{5757800}, and QoS-aware scheduling (QoS)~\cite{KOUTLIA2023102745}.
We measure all metrics as end-to-end between UE and the server at IP-layer.

\subsection{Scenario 1: Motion Control}
This scenario evaluates high-density motion control systems typical in factory automation.

The parameters for this scenario are listed in Table~\ref{tab:Parameters}.
Each UE generates bidirectional deterministic traffic with $50\unit{B}$ every $500\unit{\mu s}$ with 5QI $82$.
The service area is $50\unit{m} \times 10\unit{m} \times 10\unit{m}$ with UE speed of $36\unit{km/h}$.
We scale the number of UEs from low to high density to evaluate scheduler performance under increasing load.

\begin{figure}[t]
	\begin{center}
		\includegraphics[width=\columnwidth]{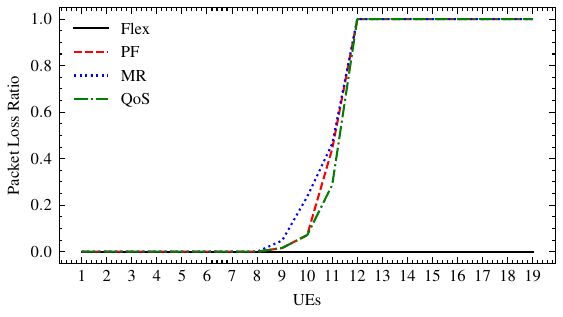}
	\end{center}
	\caption[]{PLR for different numbers of UEs in scenario 1. Lower is better. FLEX offers significantly improved performance for more than 10 UEs.}
	\label{fig:scenario1_results}
\end{figure}

Figure~\ref{fig:scenario1_results} shows the number of packets received depending on the number of UEs.
For low UE density ($<10$ UEs), FLEX matches the performance of PF, with all schedulers performing similarly.
However, beyond $10$ UEs, FLEX significantly outperforms all baseline schedulers.

The root cause of baseline scheduler failure occurs in 5G-LENA: when the number of UEs exceeds the available symbols (typically $12$ usable symbols per slot after overhead), proportional allocation calculates $w_i < 1$ for each UE, which rounds to zero symbols allocated.
This causes systematic packet drops.
FLEX's priority-based allocation correctly handles this edge case by allocating resources one by one, instead of proportionally.

Packet loss ratio (PLR) follows the same pattern, with FLEX maintaining PLR below $1\%$ even at high densities, while baseline schedulers experience dramatic increases in packet loss beyond $12$ UEs.
Latency measurements show minimal impact from FLEX's buffer estimation, significantly lower than one slot duration.
This means that most packets are scheduled within the same slot as baseline schedulers.
The still very low latency can be explained by the ingress prediction mechanism, which reduces latency for deterministic traffic.

\subsection{Scenario 2: Mobile Robots}
This scenario simulates autonomous mobile robots in warehouses.
Each UE generates bidirectional semi-deterministic traffic: $145\unit{B}$ every $25\unit{ms} \pm 7.25\unit{ms}$ (simulating clock drift) with 5QI $82$.
The service area is $1\unit{km} \times 1\unit{km} \times 10\unit{m}$ with UE speed of $25\unit{km/h}$.
Up to $80$ UEs are deployed.
The parameters for this scenario are listed in Table~\ref{tab:Parameters}.
This scenario tests FLEX's robustness to semi-deterministic traffic patterns with clock drift.
\begin{figure}[t]
	\begin{center}
		\includegraphics[width=\columnwidth]{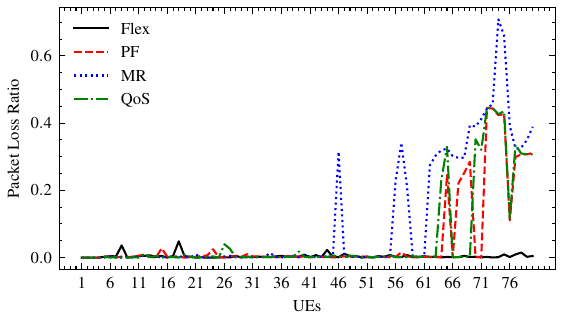}
	\end{center}
	\caption[]{PLR for different numbers of UEs in scenario 2. Lower is better. FLEX offers significantly improved performance for a high number of UEs.}
	\label{fig:scenario2_results}
\end{figure}
Throughput results mirror Scenario 1, though benefits appear at higher UE counts ($>60$ UEs) due to the longer \SI{25}{ms} packet interval reducing simultaneous transmission probability.
Since the message interval is significantly higher than in Scenario 1, the probability of more than $12$ UEs being available for transmission decreases, and existing schedulers start dropping packets later.
Figure~\ref{fig:scenario2_results} reveals that FLEX incurs a latency penalty of approximately $k2$ slots (\SI{1}{ms} for $k2=2$) compared to the PF scheduler.
This penalty stems from the semi-deterministic traffic pattern: clock drift ($\pm 25\%$) increases the CV, causing ingress prediction to be disabled for UEs currently affected by high burst interval deviation.
Without prediction, FLEX falls back to DL scheduling without estimation, which inherently adds $k2$ slots delay.
Despite this penalty, FLEX maintains superior throughput.

\subsection{Scenario 3: Heterogeneous QoS}
\begin{figure}[t]
	\begin{center}
		\subfloat[PLR per traffic direction in scenario 3 for QoS Scheduler. DL traffic is not prioritized, although it has higher priority.\label{fig:scenario3_results_qos}]{\includegraphics[width=\columnwidth]{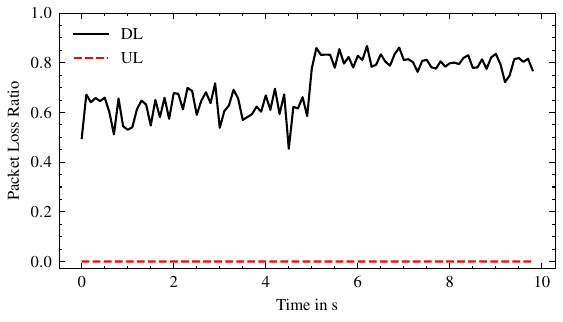}}\\
		\subfloat[PLR per traffic direction in scenario 3 for FLEX Scheduler. DL traffic is prioritized correctly.\label{fig:scenario3_results_flex}]{\includegraphics[width=\columnwidth]{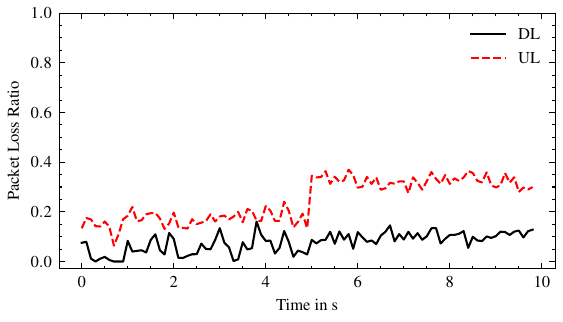}}%
	\end{center}
	\caption[]{PLR per traffic direction in scenario 3. Lower is better. DL traffic has a higher priority and DL PLR should remain lower than UL traffic PLR.}
\end{figure}

This scenario evaluates asymmetric traffic with heterogeneous QoS requirements:

The parameters for this scenario are listed in Table~\ref{tab:Parameters}.
This scenario specifically tests whether schedulers correctly prioritize high-priority DL traffic over low-priority UL traffic in asymmetric load conditions.
Scenario 3 is configured such that at the start of the simulation, the cell is already at maximum capacity.
This scenario is configured with 1 UL and 2 DL flows.
The UL flow with low priority QoS is constantly transmitting at $\SI{50}{B}$ every $50\unit{\mu s}$.
The 2 DL flows transmit $\SI{50}{B}$ every $50\unit{\mu s}$ with high QoS priority.
One of the DL flows starts delayed at \SI{5}{s} simulation time.
\begin{itemize}
	\item \textbf{Flow 1}: $4$ UEs, unidirectional UL, $50\unit{B}$ every $50\unit{\mu s}$, 5QI $83$ (priority $20$, lower)
	\item \textbf{Flow 2}: $2$ UEs, unidirectional DL, $50\unit{B}$ every $50\unit{\mu s}$, 5QI $82$ (priority $19$, higher)
	\item \textbf{Flow 3}: $2$ UEs, unidirectional DL, $50\unit{B}$ every $50\unit{\mu s}$, 5QI $82$, starts at $t=5\unit{s}$
\end{itemize}
The cell being at maximum capacity implies that some packets have to be dropped from the start.
Because DL traffic (Flows 2 and 3, priority $19$) has higher QoS priority than UL traffic (Flow 1, priority $20$), UL traffic should be dropped to satisfy DL requirements.
However, existing schedulers fail in this scenario and drop DL traffic instead, because their UL scheduler is not restricted and takes all available resources.
Figure~\ref{fig:scenario3_results_qos} shows that for the QoS scheduler, even when additional DL UEs start transmitting at $t=\SI{5}{s}$, no additional DL traffic is received.

In contrast, FLEX correctly enforces QoS priorities, as shown in Figure~\ref{fig:scenario3_results_flex}.
High-priority DL traffic receives resources preferentially over low-priority UL.
When Flow 3 begins at $t=\SI{5}{s}$, FLEX immediately reduces UL allocations to accommodate the additional high-priority DL demand.
The correct resource assignment can be attributed to the priority function combined with the buffer state estimation mechanism that restricts the UL scheduler.
Nevertheless, even the high-priority DL traffic experiences some packet loss.
The cause of the packet loss is the PF term in the priority function, which occasionally ranks UL traffic higher when the past achieved data rate drops very low.
This scenario demonstrates that FLEX is capable of fairly handling asymmetric traffic with heterogeneous QoS requirements in flexible TDD slots.

\section{Conclusion}%
\label{sec:Conclusion}
This paper presented FLEX, a novel QoS-aware scheduler for dynamic TDD resource allocation in industrial 5G networks.
FLEX addresses the critical challenge of DL starvation in flexible TDD slots by introducing DL buffer state estimation at UL scheduling time ($k2$ slots ahead), enabling fair resource allocation between traffic directions.
For industrial scenarios with deterministic traffic patterns, FLEX further exploits traffic predictability to accurately estimate future buffer states, enabling proactive resource management.

Through comprehensive simulations across three industrial scenarios, we demonstrated that FLEX achieves several key benefits:
\begin{itemize}
	\item \textbf{Superior QoS Enforcement}: FLEX is the only evaluated scheduler that correctly prioritizes high-priority DL traffic over low-priority UL traffic in asymmetric scenarios, preventing DL starvation.
	\item \textbf{Minimal Latency Impact}: For deterministic traffic, FLEX maintains latency within $1$ slot duration of established schedulers, making it suitable for URLLC applications.
	\item \textbf{Robustness}: FLEX maintains throughput advantages even with semi-deterministic traffic ($\pm 25\%$ clock drift), though with a $k2$-slot latency penalty when prediction is disabled.
\end{itemize}

The primary limitation of FLEX is the restriction of each UE to only one traffic flow in UL.
This restriction prevents the usage of 5G gateways, which could potentially be used to connect multiple sensors with a single modem.
Additionally, there might be a latency penalty applied to DL traffic (approximately $k2$ slots) when traffic becomes highly non-deterministic, as ingress prediction must be disabled.
However, industrial automation scenarios typically feature deterministic or semi-deterministic traffic patterns where this penalty is minimal or absent.

\textbf{Future Work.}
Several directions merit further investigation:
\begin{enumerate}
	\item Extension to Frequency Division Multiple Access (FDMA) scheduling strategies beyond the current TDMA implementation.
	\item Investigation of different priority functions and machine learning techniques for improved traffic prediction in semi-deterministic scenarios.
	\item Evaluation of FLEX in multi-cell deployments with inter-cell interference considerations.
\end{enumerate}

FLEX represents a significant step toward realizing the full potential of flexible TDD for industrial 5G deployments, enabling dynamic adaptation to asymmetric traffic demands while maintaining strict QoS requirements.
The scheduler's non-invasive design requires no modifications to user equipment and facilitates practical deployment in existing 5G infrastructure.

\balance
\printbibliography%
\pagebreak

\TempDisplayPreparation
\end{document}